\documentclass[twoside,11pt]{article}

\usepackage[cp1251]{inputenc}

\usepackage{amssymb}
\usepackage{amsmath}
\usepackage{amsfonts}
\usepackage{wasysym}

\usepackage{titlesec}

\titleformat*{\section}{\bfseries}

\begin{document}           

\title{\Large 
Comparative description of the evolving universe in classical and quantum geometrodynamics
}
\author{V.E. Kuzmichev, V.V. Kuzmichev\\[0.5cm]
\itshape Bogolyubov Institute for Theoretical Physics,\\
\itshape National Academy of Sciences of Ukraine, Kiev, 03680 Ukraine}

\date{}

\maketitle

\begin{abstract}
The description of the universe evolving in time according to general relativity is given in comparison with
the quantum description of the same universe in terms of semiclassical wave functions. The 
spacetime geometry is determined by the Robertson-Walker metric. It is shown that the main equation of the quantum geometrodynamics
is reduced to the non-linear Hamilton-Jacobi equation. Its non-linearity is caused by a new source of the gravitational field,
which has a purely quantum dynamical nature, and is additional to ordinary matter sources. In the semiclassical approximation,
the non-linear equation of motion is linearized and reduces to the Friedmann equation with the additional quantum source of gravity
(or anti-gravity) in the form of the stiff Zel'dovich matter. The semiclassical wave functions of the universe, in which
different types of matter-energies dominate, are obtained. As examples, the cases of the domination of radiation, barotropic fluid,
or new quantum matter-energy are discussed. The probability of the transition from the quantum state, where radiation dominates
into the state, in which barotropic fluid in the form of dust is dominant, is calculated. This probability has the same order of 
magnitude as the matter density contrast in the era of matter-radiation equality.
\end{abstract}


\section{Introduction}
The quantum theory predicts that the wave properties should be observed in the matter system which this theory describes.
The main object of the theory is the state vector (wave function) $\Psi$. The state vector satisfies the differential equations
defined in some configuration space $\Omega$. In the general case, the state vector $\Psi$ is a complex-valued function,
and without loss of generality it can always be written as $\Psi_{\alpha} = \mathcal{A}_{\alpha}\,e^{i \varphi_{\alpha}}$,
where $\mathcal{A}_{\alpha}$ and $\varphi_{\alpha}$ are real functions of generalized variables in $\Omega$,
$\alpha$ is a set of quantum numbers which characterize the state of the system with the state vector $\Psi_{\alpha}$.
In the region of $\Omega$, where the phase $\varphi_{\alpha}$ varies by a large amount on small scales, the system
under investigation can be considered as an almost classical system, in a sense that its wave properties are inessential 
and can be ignored when calculating parameters of the system \cite{Lan65}. Nevertheless, an almost classical system still has
the wave properties. They can give probabilistic character to parameters of the system which have no analogs in classical theory.
So, the overlap integral, $\langle \Psi_{\alpha'} | \Psi_{\alpha} \rangle = \int d\Omega\, \mathcal{A}^{*}_{\alpha'} \mathcal{A}_{\alpha}\,
e^{i (\varphi_{\alpha} - \varphi_{\alpha'})}$ at $\alpha' \neq \alpha$, can be nonzero because of the contribution of a subregion in
$\Omega$, where the difference $\varphi_{\alpha} - \varphi_{\alpha'}$ of two large phases is small. As a result,
the spontaneous transition (or transition under the action of an instantaneous perturbation) $\alpha \rightarrow \alpha'$
with the change of physical state of the system becomes possible. Classical and quantum descriptions of physical properties
of the same system appear here as complementary without contradiction to each other.

In the present paper, we provide the comparative description of the evolution of the universe as a whole in terms of classical
and quantum theories. As is well-known, the universe is subject to classical laws of general relativity on large spacetime scales,
whereas on small, comparable with Planck scales, it should be described from the quantum-theoretical perspective.
The questions can be posed whether the universe preserves the wave properties during its subsequent evolution and whether these properties can be discovered. 

As a working model of spacetime geometry we choose an example of a minisuperspace model with the maximally symmetric geometry
described by the Robertson-Walker metric. Despite its relative simplicity, this model may
give insights into understanding of a complete theory of quantum cosmology.
In Sect.~2, the quantum constraint equations imposed on the state vector of the universe are given. These equations are
formulated in the representation of the generalized field variables, such as the cosmic scale factor and the uniform scalar field.
The scalar field is described by some Hermitian Hamiltonian. Its mean values with respect to proper state vectors
determine the proper energy of matter in the form of barotropic fluid contained in the comoving volume (Sect.~3).
As is shown in Sect.~4, the main equation of the theory can be rewritten as the nonlinear Hamilton-Jacobi equation.
Its nonlinear part is caused by a new source of the gravitational field, which has a purely quantum dynamical nature, and 
is additional to ordinary matter sources. In Sect.~5, the classical description of the universe evolving in time according to
power and exponential laws are given in comparison with the quantum description of the same universe in the language of
wave functions in the semiclassical approximation. The semiclassical wave functions of the universe, in which
different types of matter-energies dominate, are obtained. As examples, the cases of the domination of radiation or barotropic fluid
are discussed. The case of domination of new quantum matter-energy is special. It is shown that its energy density is negative, while its
equation of state coincides with the equation of state of the stiff Zel'dovich matter. Such an energy density dominates
in the sub-Planck region. Here the wave function is constant, and the semiclassical equation of motion has an allowed trajectory in
imaginary time. The fact that the wave function is non-vanishing near the initial singularity point means that in this region there is
some source which provides the origin of the universe with a finite nucleation rate \cite{Kuz09,Kuz10} (cf. Refs.~\cite{Har83,Gib90,Bou98}). 
In Sect.~6, as a supplement, the transition probability of the universe from the state, where radiation dominates, into the state, 
in which barotropic fluid in the form of dust is dominant, is calculated.

Throughout the paper, unless otherwise specified, the modified Planck system of units is used. 
As a result, all quantities in the equations become dimensionless.
The length $l_{P} = \sqrt{2 G \hbar / (3 \pi c^{3})}$ is taken as a unit of length and the $\rho_{P} = 3 c^{4} / (8 \pi G l_{P}^{2})$ 
is used as a unit of energy density and pressure. The mass-energy is measured in units of Planck mass,
$m_{P} c^{2} = \hbar c / l_{P}$. The proper time $\tau$ is
taken in units of $l_{P}$. The time parameter (conformal time) $T$ is expressed in radians. 
The scalar field is taken in $\phi_{P} = \sqrt{3 c^{4} / (8 \pi G)}$. Here $G$ is Newton's gravitational constant.

\section{Quantum constraint equations}
We consider a canonical approach to the quantization of general relativity, in particular quantum geometrodynamics. 
It is well-known that approaches of this type share a common problem known as 
the `problem of time' (see, e.g. Refs.~\cite{Ish,Kuch92}). The apparent manifestation of this problem is 
the absence of explicit time parameter in the Wheeler-DeWitt equation considered as the main 
dynamical equation of the theory.

The way to solve this problem could be rewriting the classical 
constraint equations to obtain a Schr\"{o}dinger-type equation, as a preliminary to quantization.
Since it was proved that general relativity could not be viewed as a parametrized field theory 
\cite{Tor}, a concept of matter clocks and reference fluids was proposed \cite{Kuch92}. 
One approach in this direction goes back to DeWitt, who studied a coupling of clocks to an elastic 
media \cite{DeW}. It can be shown that perfect fluids are a special case of DeWitt's 
relativistic elastic media and that in the velocity-potential formalism for perfect 
fluids the thermasy can be connected to a clock variable with DeWitt-type 
coupling \cite{Bro}. Kijowski et al. have proposed a physical interpretation for the thermasy 
as a ``material time'' relating it to ``proper time retardation'' due to the chaotic motion of the particles 
of the fluid moving chaotically around the flow lines \cite{Kij,JKi}. 
The perfect fluid plays a crucial role in the formalism as the ``reference fluid'', 
whose particles identify spacetime points, and clocks carried with them identify  
instants of time (cf. Ref.~\cite{Kuch92}).

We are working in the framework of the velocity-potential version of perfect-fluid hydrodynamics 
formulated by Seliger and Whitham \cite{SeW}, and generalized by Schutz \cite{Sch70} (see also \cite{Br92}). 
In this formalism, the perfect fluid four-velocity $U^{\nu}$ is written as a combination of (at least) five scalar 
fields (potentials) and their gradients. An independent physical interpretation can be given
to each potential. So, one potential appears to be the thermasy $\Theta$ introduced earlier by 
van Dantzig \cite{Dan}; another is the entropy $s$, and the other one is the potential 
$\tilde \lambda$ for the specific free energy and so on. In the general case, all five potentials 
are needed in order to describe a relativistic perfect fluid. In the cosmological model with the 
Robertson-Walker metric and irrotational flows of a perfect fluid, the number of potentials reduces, 
and the perfect fluid can be described by three potentials: $\Theta$, $s$, and $\tilde \lambda$.

Let us consider the homogeneous and isotropic cosmological model and write the
Robertson-Walker line element in the form
\begin{equation}\label{01}
     ds^{2} = a^{2} [dT^{2} - d\Omega_{3}^{2}],
\end{equation}
where $a$ is the cosmic scale factor which is a function of time, $T$ is the time variable connected with the proper time $\tau$ 
by the differential equation $d\tau = a dT$, $T$ is the ``arc-parameter measure of time'': during the interval $d\tau$, a photon
moving on a hypersphere of radius $a(\tau)$ covers an arc $dT$ measured in radians \cite{MTW}.
$d\Omega_{3}^{2}$ is a line element on a unit three-sphere. Following the ADM formalism \cite{Dir58,ADM},
one can extract the so-called lapse function $N$, that specifies the time reference scale, from the total differential $dT$: $dT = N d \eta$,
where $\eta$ is the ``arc time'' which coincides with $T$ for $N = 1$ (cf. Refs. \cite{MTW,Lan2}). In the general case, the function $N$
plays the role of the Lagrange multiplier in the Hamiltonian formalism and it should be taken into account in an appropriate way.

Having in mind mentioned above, we can write down the Hamiltonian for the cosmological system (universe) 
in the form \cite{Kuz08,Kuz13,Kuz15}
\begin{eqnarray}\label{02}
    H & = & \frac{N}{2} \left\{-\,\pi_{a}^{2} - a^{2} + a^{4} [\rho_{\phi} + \rho_{\gamma}]\right\} \nonumber \\ 
& + & \lambda_{1}\left\{\pi_{\Theta} - \frac{1}{2}\,a^{3} \rho_{0} s\right\}
+ \lambda_{2}\left\{\pi_{\tilde{\lambda}} + \frac{1}{2}\,a^{3} \rho_{0} \right\},
\end{eqnarray}
where $\pi_{a},\, \pi_{\Theta},\, \pi_{\tilde{\lambda}}$ are the momenta 
canonically conjugate with the variables $a,\, \Theta,\, \tilde{\lambda}$, $\rho_{\phi}$ is the energy density of matter (the field $\phi$),
$\rho_{\gamma} (\rho_{0}, s)$ is the energy density of a perfect fluid, which defines a material reference frame \cite{DeW,Bro},
and it is a function of the density of the rest mass $\rho_{0}$ and the specific entropy $s$; 
the $\Theta$ is the thermasy which defines the temperature, $\mathcal{T} = \Theta_{,\nu} U^{\nu}$; the $U^{\nu}$ 
is the four-velocity; the
$\tilde{\lambda}$ is the potential for the specific Gibbs free energy $\mathcal{F}$ taken with an inverse sign, 
$\mathcal{F} = - \tilde{\lambda}_{,\nu} U^{\nu}$. 
The $N$, $\lambda_{1}$, and $\lambda_{2}$ are the Lagrange multipliers.

The Hamiltonian (\ref{02}) is a linear combination of constraints (expressions in braces) and thus weakly vanishes, $H \approx 0$.
The variations of the Hamiltonian with respect to $N$, $\lambda_{1}$, and $\lambda_{2}$ give three constraint equations,
\begin{equation}\label{03}
-\,\pi_{a}^{2} - a^{2} + a^{4} [\rho_{\phi} + \rho_{\gamma}] \approx 0, \quad \pi_{\Theta} - \frac{1}{2}\,a^{3} \rho_{0} s \approx 0, 
\quad \pi_{\tilde{\lambda}} + \frac{1}{2}\,a^{3} \rho_{0} \approx 0.
\end{equation}
From the conservation of these constraints in time, it follows that the number of particles of a perfect fluid in the proper 
volume\footnote{This volume is equal to $2 \pi^{2} a^{3}$, where $a$ is taken in units of length.}
$\frac{1}{2} a^{3}$ and the specific entropy conserve: $E_{0} \equiv \frac{1}{2} a^{3} \rho_{0} = \mbox{const}$, $s =  \mbox{const}$.
Taking into account these conservation laws and vanishing of the momenta conjugate with the variables $\rho_{0}$ and $s$,
$\pi_{\rho} = 0$ and $\pi_{s} = 0$,
one can discard degrees of freedom corresponding to these variables, and convert the second-class constraints into
first-class constraints in accordance with Dirac's proposal \cite{Dir64}.

It is convenient to choose the perfect fluid with the density $\rho_{\gamma}$ in the form of relativistic matter (radiation).
Then, in Eq. (\ref{03}) one can put $a^{4} \rho_{\gamma} \equiv E = \mbox{const}$. The matter field with the energy density 
$\rho_{\phi}$ and pressure $p_{\phi}$ can be taken for definiteness in the form of a uniform scalar field $\phi$,
\begin{equation}\label{04}
    \rho_{\phi} = \frac{2}{a^{6}}\,\pi_{\phi}^{2} + V(\phi), \quad p_{\phi} = \frac{2}{a^{6}}\,\pi_{\phi}^{2} - V(\phi),
\end{equation}
where $V(\phi)$ is the potential of this field, $\pi_{\phi}$ is the momentum conjugate with $\phi$.
After averaging with respect to appropriate quantum states, the scalar field turns into the effective matter fluid 
(see Ref. \cite{Kuz13}, and below).

In quantum theory, first-class constraint equations (\ref{03}) become constraints on the state vector $\Psi$ \cite{Dir64}
and, in this way, define the space of physical states, which can be turned into a Hilbert space (cf. Ref. \cite{Kuc91}).
Passing from classical variables in Eqs. (\ref{02})-(\ref{04}) to corresponding operators, using the conservation laws, and
introducing the non-coordinate co-frame
\begin{equation}\label{06}
    h d\tau = s d\Theta - d\tilde{\lambda}, \quad h dy = s d\Theta + d\tilde{\lambda},
\end{equation}
where $h = \frac{\rho_{\gamma} + p_{\gamma}}{\rho_{0}}$ is the specific enthalpy, $p_{\gamma}$ is the pressure of radiation,
and $y$ is a supplementary variable, we obtain
\begin{equation}\label{1}
\left(-i \partial_{T} - \frac{2}{3} E \right) \Psi = 0, \quad \partial_{y} \Psi = 0,
\end{equation}
\begin{equation}\label{2}
\left(- \partial_{a}^{2} + a^{2} - 2 a \hat{H}_{\phi} - E \right)\Psi = 0,
\end{equation}
where
\begin{equation}\label{08}
\hat{H}_{\phi} = \frac{1}{2} a^{3} \hat{\rho}_{\phi}
\end{equation}
is the operator of Hamiltonian of the scalar field $\phi$ which is Hermitian, and the operator $\hat{\rho}_{\phi}$ is described by Eq. (\ref{04}) 
with $\pi_{\phi} = -i \partial_{\phi}$. Equation (\ref{2}) is the Wheeler-DeWitt equation of a minisuperspace model for the universe filled
with a scalar field and radiation, when the state vector $\Psi$ does not depend on time. In the approach under consideration, the 
dependence of the state vector $\Psi$ on time is determined from Eq.~ (\ref{1}).

The quantum constraints (\ref{1}) and (\ref{2}) can be rewritten in the form of the time-dependent Schr\"{o}dinger-type equation
\begin{equation}\label{3}
- i \partial_{T} \Psi = \frac{2}{3} \mathcal{H} \Psi,
\end{equation}
where
\begin{equation}\label{4}
\mathcal{H} = - \partial_{a}^{2} + a^{2} - 2 a \hat{H}_{\phi}.
\end{equation}
The minus sign before the partial derivative $\partial_{T}$ is stipulated by the specific character of the cosmological problem,
namely that the classical momentum conjugate to the variable $a$ is defined with the minus sign \cite{Lin90,Kuz02} (see below).

The partial solution of Eqs.~(\ref{1}) and (\ref{2}) has a form
\begin{equation}\label{5}
\Psi (T) = e^{i \frac{2}{3} E (T - T_{0})} \Psi (T_{0}),
\end{equation}
where the vector $\Psi (T_{0}) \equiv \langle a, \phi | \psi \rangle$ satisfies the stationary equation
\begin{equation}\label{6}
\mathcal{H} | \psi \rangle = E | \psi \rangle.
\end{equation}
From the condition 
\begin{equation}\label{7}
0 = \frac{d}{dT} \int D[a, \phi]\, |\Psi |^{2} = - i \frac{2}{3} \int D[a, \phi]\, \Psi^{*} \left[\mathcal{H}^{\dag} - \mathcal{H} \right] \Psi,
\end{equation}
where $D[a, \phi]$ is the measure of integration with respect to the fields $a$ and $\phi$ chosen in an appropriate way,
it follows that the operator (\ref{4}) is Hermitian: $\mathcal{H} = \mathcal{H}^{\dag}$.

\section{Barotropic fluid}
The Hamiltonian of matter $\hat{H}_{\phi}$ can be diagonalized by means of some state vectors $\langle x | u_{k} \rangle$
in the representation of generalized field variable $x = x (a, \phi)$ with the measure of integration
$D[a, \phi] = da\,dx$ in Eq.~(\ref{7}).

Assuming that the states $| u_{k} \rangle$ are orthonormalized, $\langle u_{k} | u_{k'} \rangle = \delta_{k k'}$,
we obtain the equation
\begin{equation}\label{8}
\langle u_{k} | \hat{H}_{\phi} | u_{k'} \rangle = M_{k}(a)\,\delta_{k k'},
\end{equation}
which determines the proper energy $M_{k}(a) = \frac{1}{2} a^{3} \rho_{m}$ of a substance (barotropic fluid) in discrete and/or
continuous $k$th state in the comoving volume $\frac{1}{2} a^{3}$ with the energy density 
$\rho_{m} = \langle u_{k} | \hat{\rho}_{\phi}| u_{k} \rangle$ and pressure
$p_{m} = \langle u_{k} | \hat{p}_{\phi}| u_{k} \rangle = w_{m}(a) \rho_{m}$, where 
\begin{equation}\label{9}
w_{m}(a) = - \frac{1}{3} \frac{d \ln M_{k} (a)}{d \ln a},
\end{equation}
is the equation of state parameter.

In order to obtain the general form of the proper energy $M_{k}(a)$, let us suppose that the interaction of the uniform scalar field
is described by the potential $V(\phi) = \lambda_{\alpha} \phi^{\alpha}$, where $\lambda_{\alpha}$ is the coupling constant and
$\alpha$ takes arbitrary non-negative values, $\alpha \geq 0$. Then we find
\begin{equation}\label{10}
M_{k}(a) = \epsilon_{k} \left(\frac{\lambda_{\alpha}}{2} \right)^{\frac{2}{2 + \alpha}} a^{\frac{3 (2 - \alpha)}{2 + \alpha}},
\end{equation}
where $\epsilon_{k}$ is an eigenvalue of the equation $(-\partial_{x}^{2} + x^{\alpha} - \epsilon_{k}) |u_{k} \rangle = 0$,
and $x = \left(\frac{\lambda_{\alpha} a^{6}}{2} \right)^{\frac{1}{2 + \alpha}} \phi$ is the rescaled matter scalar field.
The equation of state parameter in such a model does not depend on $a$ and has a simple form 
\begin{equation*}
w_{m}(a) = \frac{\alpha - 2}{\alpha + 2}.
\end{equation*}
It describes the barotropic fluid in all possible states. In the case of the model $\phi^{0}$, the field 
$\phi$ averaged over its quantum states reproduces vacuum (dark energy) in the $k$th state with the density
$\rho_{m} = \lambda_{0} \epsilon_{k}$ and the function $\langle x | u_{k} \rangle$ in the plane wave form $e^{ikx}$
with the wave vector $k = \pm \sqrt{\epsilon_{k} - 1}$. The model $\phi^{1}$ describes the strings in the 
$k$th state with the energy density $\rho_{m} = \left(\frac{\lambda_{1}}{2} \right)^{2/3} \frac{2 \epsilon_{k}}{a^{2}}$,
where $\epsilon_{k} \lessgtr 0$ and $| u_{k} \rangle$ is the Airy function. In the model $\phi^{2}$, 
the scalar field, after averaging over quantum states, turns into dust with the total mass $M_{k} = \sqrt{2 \lambda_{2}}(k + \frac{1}{2})$,
where $k$ is the number of dust particles, and the density $\rho_{m} = \frac{2 M_{k}}{a^{3}}$. The model $\phi^{4}$ 
leads to the relativistic matter with the energy density $\rho_{m} = \left(\frac{\lambda_{4}}{2} \right)^{1/3} \frac{2 \epsilon_{k}}{a^{4}}$,
where $\epsilon_{k} < \infty$ and $| u_{k} \rangle$ has the asymptotics in the form of cylindrical function. In the case
$\alpha = \infty$, the field $\phi$, averaged over the states $| u_{k} \rangle$, reduces to the stiff Zel'dovich matter with
the density $\rho_{m} = \frac{2 \epsilon_{k}}{a^{6}}$.

\section{Non-linear Hamilton-Jacobi equation}
Assuming that the set of vectors $| u_{k} \rangle$ is complete, $\sum_{k} | u_{k} \rangle \langle u_{k}| = 1$, the solution of
Eq.~(\ref{6}) can be represented in the form of the superposition of states of the universe with the substance in the $k$th state
in any form described above. We have
\begin{equation}\label{11}
| \psi \rangle = \sum_{k} | u_{k} \rangle \langle u_{k}| \psi \rangle,
\end{equation}
where the wave function $f(a) \equiv \langle u_{k}| \psi \rangle$ satisfies the equation
\begin{equation}\label{12}
\left[- \partial_{a}^{2} + a^{2} - 2 a M(a) \right] f = E f.
\end{equation}
The index $k$ is omitted here and below, since in what follows we consider the universe with the proper energy of the substance
in a specific $k$th state, $M_{k}(a) \equiv M(a)$. Because the operator (\ref{4}) is Hermitian, it follows that the operator
on the left-hand side of Eq.~(\ref{12}) is Hermitian as well. This equation determines the wave function corresponding to the 
particular eigenvalue $E$. Depending on the form of $M(a)$, the constant $E$ can take the values lying in a
discrete or continuous spectrum of the states of the proper energy of radiation $M_{\gamma} = \frac{1}{2} a^{3} \rho_{\gamma}$
with the energy density $\rho_{\gamma} = \frac{E}{a^{4}}$ and pressure $p_{\gamma} = \frac{1}{3} \rho_{\gamma}$.
Thus, the value of the constant $E$ is determined through the quantum numbers enumerating the states of the substance and radiation.

We look for the solution of Eq.~(\ref{12}) in the form of the wave propagating along the $a$ direction
\begin{equation}\label{13}
f(a) = A e^{i S(a)},
\end{equation}
where $A$ is the normalizing constant, and the phase $S(a)$ is a complex function
\begin{equation}\label{14}
S(a) = S_{R}(a) + i S_{I}(a)
\end{equation}
(with $S_{R}$ and $S_{I}$ real). 
In the general case, the solution of Eq.~(\ref{12}) is the superposition of the wave function $f(a)$ and its complex conjugate
$f^{*}(a)$. Substituting Eq.~(\ref{13}) into Eq.~(\ref{12}), we find that $S_{R}(a)$ satisfies the non-linear equation
\begin{equation}\label{15}
\left(\partial_{a} S_{R} \right)^{2} + a^{2} - 2 a M(a)  - E = Q(a),
\end{equation}
where the function
\begin{equation}\label{16}
Q(a) = \frac{3}{4} \left(\frac{\partial_{a}^{2} S_{R}}{\partial_{a} S_{R}} \right)^{2} - \frac{1}{2} \frac{\partial_{a}^{3} S_{R}}{\partial_{a} S_{R}}
\end{equation}
describes the new source of the gravitational field with the energy density
\begin{equation}\label{17}
\rho_{Q} = \frac{Q(a)}{a^{4}},
\end{equation}
which is additional to the ordinary matter (substance and radiation). This source has the quantum dynamical nature. It emerges
as a result of expansion (or contraction) of the universe as a whole. The equation of state of quantum source of matter-energy
has a form
\begin{equation}\label{18}
p_{Q} = w_{Q}(a) \rho_{Q},
\end{equation}
where $p_{Q}$ is the pressure, and
\begin{equation}\label{19}
w_{Q}(a) = \frac{1}{3} \left(1 -  \frac{d \ln Q(a)}{d \ln a} \right)
\end{equation}
is the equation of state parameter. The first term in Eq.~(\ref{19}) takes into account the correction for relativity.
The additional source of matter-energy has a proper energy $M_{Q} = \frac{1}{2} a^{3} \rho_{Q} = \frac{Q}{2 a}$
contained in the volume $\frac{1}{2} a^{3}$.

The derivative of the imaginary part of the phase $S(a)$ is
\begin{equation}\label{20}
\partial_{a} S_{I} = \frac{1}{2} \frac{\partial_{a}^{2} S_{R}}{\partial_{a} S_{R}}.
\end{equation}
In order to clarify the physical meaning of the quantum correction in Eq.~(\ref{15}), we rewrite the source function $Q(a)$
as
\begin{equation}\label{21}
Q(a) = \left(\partial_{a} S_{I} \right)^{2} - \partial_{a}^{2} S_{I},
\end{equation}
where Eq.~(\ref{20}) was used. It follows from here that if the imaginary part $S_{I}$ of the phase (\ref{14})
is a slowly varying function of $a$, then one can set $Q(a) \approx 0$. In this case, Eq.~(\ref{15}) becomes 
the Hamilton-Jacobi equation for the classical action $S_{cl}$ and the wave function (\ref{13}) takes the form
\begin{equation}\label{22}
f(a) = \mbox{const}\,e^{i S_{cl}}.
\end{equation}
It describes the quantum universe in the semiclassical approximation, when $S_{R} = S_{cl}$.
In classical mechanics, the momentum is equal to the first derivative of the action with respect to the generalized coordinate.
In general relativity, Hamilton's equations of motion lead to \cite{Kuz02}
\begin{equation}\label{23}
\partial_{a} S_{cl} = - \frac{d a}{d T} = - a \dot{a},
\end{equation}
where we denote $\dot{a} = \frac{d a}{d \tau}$. Equations (\ref{22}) and (\ref{23}) describe the same universe, but from the
different point of view, namely as the quantum system in the approximation $S_{R} = S_{cl}$ or as the classical object obeying the laws
of general relativity.

To determine the physical meaning of the derivative $\partial_{a} S_{R}$, we calculate the probability flux density
for the universe to be a hypersurface with radius $a$ in a four-dimensional space. From Eq.~(\ref{15}), it follows that
the probability flux density is described by the expression
\begin{equation}\label{24}
J_{a} = \frac{1}{2 i} \left(f^{*}\partial_{a} f - f \partial_{a} f^{*} \right).
\end{equation}
Taking into account Eq.~(\ref{13}), we find
\begin{equation}\label{25}
J_{a} = |f|^{2}\, \partial_{a} S_{R},
\end{equation}
where $|f|^{2} = |A|^{2} e^{- 2 S_{I}}$ is the probability density of the universe to have the scale factor $a$. Equation (\ref{25})
shows that the wave function (\ref{13}) describes the expansion of the universe as a whole with the generalized momentum 
$\partial_{a} S_{R}$. The generalized action $S_{R}$ is the solution of the non-linear Hamilton-Jacobi equation (\ref{15}).

One can make sure that the energy density (\ref{17}) is the quantum correction to the energy density of the substance and
radiation by rewriting Eq.~(\ref{15}) in dimensional physical units
\begin{equation}\label{26}
\left(\partial_{a} S_{R} \right)^{2} + \left(\frac{3 \pi c^{3}}{2 G} \right)^{2} a^{2} \left[1 - \frac{8 \pi G}{3 c^{4}}  a^{2}
\left(\rho_{m} + \rho_{\gamma} + \rho_{Q} \right) \right] = 0,
\end{equation}
where
\begin{equation}\label{27}
\rho_{m} = \frac{M(a)}{2 \pi^{2} a^{3}}, \quad \rho_{\gamma} = \frac{E}{a^{4}}, \quad 
\rho_{Q} = \frac{\hbar^{2}}{6 \pi^{3}} \frac{G}{c^{2}} \frac{Q(a)}{a^{4}}
\end{equation}
are the energy densities of the substance, radiation, and quantum addition measured in GeV/cm$^{3}$. Here 
$a$ is taken in cm, $M(a)$ in GeV, $E$ in GeV cm ($\hbar c$), $S_{R}$ in GeV s ($\hbar$), whereas $Q$ is in cm$^{-2}$ 
and it has the same form as in Eq.~(\ref{16}). From Eq.~(\ref{26}), it follows that the quantum correction is 
proportional to $\hbar^{2}$. In the formal limit $\hbar \rightarrow 0$, Eq.~(\ref{26}) turns into the Hamilton-Jacobi equation
of general relativity. A more rigorous approach requires the change to dimensionless variables which do not contain
dimensional fundamental constants $G$, $c$, and $\hbar$. The impact of
the quantum correction $Q(a)$ on the dynamics of the universe as a whole is determined by how quickly the amplitude
$A\,e^{- S_{I}(a)}$ of the wave function (\ref{13}) changes with $a$. In Eq.~(\ref{15}), this impact depends on how
its terms behave with increasing (decreasing) of $a$. In the models, in which Eq.~(\ref{12}) can be integrated exactly, 
Eq.~(\ref{15}) also admits a solution in an analytical form \cite{Kuz08,Kuz13}.

\section{Classical-quantum correspondence}
In order to determine the correspondence between classical and quantum description of the same physical system,
we consider the specific model of the universe. Let us assume that the universe evolves in time $\tau$ 
with a power-law scale factor
\begin{equation}\label{28}
a = \beta\, \tau^{\alpha},
\end{equation}
where $\alpha$ and $\beta$ are some arbitrary constants\footnote{Here the constant $\alpha$ differs from the
parameter $\alpha$ in Eq.~(\ref{10}). We use the same letter to emphasize that, in both cases, different types of matter-energy
correspond to different values of $\alpha$.}. Then the generalized momentum in the semiclassical approximation is equal to
\begin{equation}\label{29}
\partial_{a} S_{R} = - \alpha\,\beta^{\frac{1}{\alpha}} a^{\frac{2 \alpha - 1}{\alpha}},
\end{equation}
and the quantum source function (\ref{16}) takes the form
\begin{equation}\label{30}
Q(a) = \frac{\gamma_{\alpha}}{a^{2}},
\end{equation}
where the numerator
\begin{equation}\label{31}
\gamma_{\alpha} = \frac{(2\alpha - 1) (4 \alpha - 1)}{4 \alpha^{2}}
\end{equation}
does not depend on $a$. Universal dependence of the source function (\ref{30}) on the scale factor allows one to
find the equation of state for quantum matter-energy for any values of the parameters $\alpha$ and $\beta$.
So, it follows from Eq.~(\ref{19}) and (\ref{30}) that the equation of state parameter (\ref{19}) is $w_{Q} = 1$, and
the energy density decreases with the increase of $a$ according to the law
\begin{equation}\label{32}
\rho_{Q} = \frac{\gamma_{\alpha}}{a^{6}}.
\end{equation}
As was mentioned in Sect.~3, the stiff Zel'dovich matter has such a density. The energy density of this quantum matter
can be negative ($\frac{1}{4} < \alpha < \frac{1}{2}$), positive ($\alpha > \frac{1}{2}$, and $\alpha < \frac{1}{4}$), 
or vanish ($\alpha = \frac{1}{2}$, and $\alpha = \frac{1}{4}$).

From the point of view of quantum theory, the expansion of the universe with the scale factor (\ref{28}) is described by
the semiclassical wave function
\begin{equation}\label{33}
f_{\alpha} (a) = A_{\alpha}\,a^{-\frac{2 \alpha - 1}{2 \alpha}}\, \exp \left \{-i \frac{\alpha^{2} \beta^{\frac{1}{\alpha}}}{3 \alpha - 1}\, 
a^{\frac{3 \alpha - 1}{\alpha}} \right \} \quad \mbox{at} \ \alpha \neq \frac{1}{3},
\end{equation}
and
\begin{equation}\label{34}
f_{\frac{1}{3}} (a) = A_{\frac{1}{3}}\, a^{\frac{1}{2} - i \frac{\beta^{3}}{3}}.
\end{equation}
The complex conjugate function $f_{\alpha}^{*} (a)$ corresponds to the wave propagating towards the  
initial cosmological singularity point, $a = 0$, and describes the contracting universe.

In the limit of infinitely large values of $\alpha$, we have
\begin{equation}\label{35}
\lim_{\alpha \rightarrow \infty} \gamma_{\alpha} = 2,
\end{equation}
and the quantum correction in Eq.~(\ref{15}) equals
\begin{equation}\label{36}
Q(a) = \frac{2}{a^{2}}.
\end{equation}
The same additional term is produced by the universe expanding exponentially. Really, setting
\begin{equation}\label{37}
a = a(0)\,e^{\sqrt{\rho_{v}} \tau},
\end{equation}
where $\rho_{v}$ is some constant (e.g. $\rho_{v} = \frac{\Lambda}{3}$, where $\Lambda$ is the cosmological constant),
we calculate the momentum of the universe
\begin{equation}\label{38}
\partial_{a} S_{R} = - \sqrt{\rho_{v}}\, a^{2}.
\end{equation}
Substituting (\ref{38}) into Eq.~(\ref{16}), we obtain the quantum correction in the form (\ref{36}).
The exponentially expanding universe is described by the semiclassical wave function
\begin{equation}\label{39}
f_{\infty} (a) = A_{\infty}\, \frac{1}{a}\, e^{-i \frac{\sqrt{\rho_{v}}}{3}\,a^{3}}.
\end{equation}
At infinity this function vanishes, but it diverges at the point $a = 0$. However, this point is inaccessible, since $\tau \geq 0$ 
in Eq.~(\ref{37}). The function (\ref{39}) can be normalized to a constant. The probability flux density (\ref{25})
for the wave function (\ref{39}) is 
\begin{equation}\label{40}
J_{\infty} = - \sqrt{\rho_{v}}\,|A_{\infty}|^{2}.
\end{equation}

The approximation (\ref{30}) linearizes Eq.~(\ref{15}), and it can be considered as the Hamilton-Jacobi equation
for the generalized action $S_{R}$ with the additional source of the gravitational field with the energy density (\ref{32}).
Using Eq.~(\ref{29}), this equation can be easily reduced to the Friedmann equation for the Hubble expansion rate
$H = \frac{\dot{a}}{a}$. In the semiclassical approximation, $\partial_{a} S_{R} = - a \dot{a}$, we have
\begin{equation}\label{41}
H^{2} = \frac{2 M(a)}{a^{3}} + \frac{E}{a^{4}} + \frac{\gamma_{\alpha}}{a^{6}} - \frac{1}{a^{2}}.
\end{equation}
In addition to the energy density of the substance ($\sim M\, a^{-3}$) and radiation ($\sim a^{-4}$), this equation
contains the energy density of the quantum source ($\sim a^{-6}$).

Let us find out what restriction on the solutions of Eqs.~(\ref{15}) and (\ref{41}) is imposed by Eq.~(\ref{28}).
Since it is assumed that the universe evolves according to the law (\ref{28}) with a given $\alpha$, then,
according to the standard model,  it means that the approximation of a single component domination in the total energy density
of matter energy $\rho$ is used. It has a form \cite{PDG,Kol90}
\begin{equation}\label{42}
\rho \sim \frac{1}{a^{2/\alpha}} \quad \mbox{for} \ a \sim \tau^{\alpha}.
\end{equation}
Substitution of Eqs.~(\ref{29}) and (\ref{30}) into Eq.~(\ref{15}) gives the condition
\begin{equation}\label{43}
\alpha^{2}\,\beta^{\frac{2}{\alpha}}\,a^{\frac{2}{\alpha} (2 \alpha - 1)} + a^{2} - 2\, a\, M(a) - E = \gamma_{\alpha}\, a^{-2}.
\end{equation}

Let $\alpha = \frac{1}{2}$, then $\gamma_{\frac{1}{2}} = 0$, and Eq.~(\ref{43}) is equivalent to
\begin{equation}\label{44}
\frac{2 M(a)}{a^{3}} + \frac{E}{a^{4}} - \frac{1}{a^{2}} = \frac{\beta^{4}}{4} \frac{1}{a^{4}} .
\end{equation}
Then from Eq.~(\ref{41}), we obtain
\begin{equation}\label{45}
H^{2} =  \left( \frac{\beta^{2}}{2 a^{2}} \right)^{2} \quad \mbox{or} \quad H = \frac{1}{2 \tau},
\end{equation}
if Eq.~(\ref{28}) is used. This equation describes the spatially flat universe, in which radiation dominates. Hence, the model 
(\ref{28}) does not contradict Eq.~(\ref{41}). Such a universe expands with constant momentum
\begin{equation}\label{46}
\partial_{a} S_{R} = - \frac{\beta^{2}}{2},
\end{equation}
and, according to Eq.~(\ref{16}), an additional quantum source of energy is not generated, $Q(a) = 0$. The quantum properties 
of such a universe are described by the wave function
\begin{equation}\label{47}
f_{\frac{1}{2}} (a) = A_{\frac{1}{2}}\, e^{-i \sqrt{E} a} = A_{\frac{1}{2}}\, e^{-i H a^{3}}.
\end{equation}
This function is a solution to Eq.~(\ref{12}) in which the terms $a^{2}$ and $2 a M(a)$ are omitted, i.e. the quantum universe is
spatially flat and contains nothing but radiation. 
Comparing Eq.~(\ref{47}) with Eq.~(\ref{33}) at $\alpha = \frac{1}{2}$, we find the constant $\beta$ of Eq.~(\ref{28})
\begin{equation}\label{48}
\beta = \left(2 \sqrt{E} \right)^{1/2}.
\end{equation}
This expression coincides with the one that can be obtained directly from the solution of Eq.~(\ref{41}) in the approximation specified above.

The wave function of the continuous state $f_{\frac{1}{2}} (a) \equiv f_{\frac{1}{2}} (a; E)$ (\ref{47}) can be normalized to a delta-function
\begin{equation}\label{471}
\langle f_{\frac{1}{2}} (E) | f_{\frac{1}{2}} (E') \rangle = \delta (E - E').
\end{equation}
This condition determines the normalizing constant $A_{\frac{1}{2}}$ (up to inessential phase factor),
\begin{equation}\label{472}
|A_{\frac{1}{2}}|^{2} = \frac{1}{2 \sqrt{E}}.
\end{equation}
According to Eqs.~(\ref{46}) and (\ref{48}), the probability flux density (\ref{25}) for the universe with the wave function (\ref{47}) 
and the amplitude from Eq.~(\ref{472}) does not depend on $a$ and $E$, and it equals
\begin{equation}\label{473}
J_{\frac{1}{2}} = - \frac{1}{2}.
\end{equation}
In this case, the conservation law is fulfilled, $\partial_{a} J_{\frac{1}{2}} = 0$. The minus sign in Eq.~(\ref{473}) shows that
the matter flux is directed away from the observer, i.e. matter objects (galaxies) move away from the observer demonstrating
the effect of expansion of the universe.

Using the wave packet (proper differential)
\begin{equation}\label{474}
\bar{f}_{\frac{1}{2}} (a; E) = \int_{E - \delta}^{E + \delta} dE' f_{\frac{1}{2}} (a; E')
\end{equation}
with the width $2 \delta \ll 1$, $\delta > 0$, the wave function $f_{\frac{1}{2}} (a; E)$ can be normalized to unity as follows:
\begin{equation}\label{475}
\langle f_{\frac{1}{2}} (E) | \bar{f}_{\frac{1}{2}} (E) \rangle = 1.
\end{equation}
From Eqs.~(\ref{471}) and (\ref{474}), it follows that the integral (\ref{475}) does not depend on $\delta$. The smaller is $\delta$,
the wave packet (\ref{474}) reproduces the wave function (\ref{47}) with greater accuracy. The wave function (\ref{47}) with
the amplitude (\ref{472}) can be interpreted as a part of the de Broglie wave propagating along $a$ with the momentum 
$\partial_{a} S_{R} = - \sqrt{E}$.

In the case $\alpha = \frac{2}{3}$, we have $\gamma_{\frac{2}{3}} = \frac{5}{16}$, and the quantum source function (\ref{16}) is
\begin{equation}\label{49}
Q (a) = \frac{5}{16} \frac{1}{a^{2}}.
\end{equation}
Quantum source is characterized by the positive energy density $\rho_{Q}$ (\ref{17}). The condition (\ref{43}) is written as
\begin{equation}\label{50}
\frac{2 M(a)}{a^{3}} + \frac{E}{a^{4}} + \frac{5}{16} \frac{1}{a^{6}} - \frac{1}{a^{2}} = \frac{4}{9} \left( \frac{\beta}{a} \right)^{3},
\end{equation}
and Eq.~(\ref{41}) gives the Hubble expansion rate
\begin{equation}\label{51}
H^{2} =  \frac{4}{9} \left( \frac{\beta}{a} \right)^{3} \quad \mbox{or} \quad H = \frac{2}{3 \tau}.
\end{equation}
The latter equation describes the spatially flat universe, where non-relativistic matter with the mass $M (a) \equiv M = const$
dominates. In this case, the constant $\beta$ is determined by Eqs.~(\ref{51}) and (\ref{41}), in which domination of matter is
taken into account
\begin{equation}\label{52}
\beta = \left(\frac{3}{2} \sqrt{2 M} \right)^{2/3}.
\end{equation}
The universe expands as a whole with the momentum
\begin{equation}\label{53}
\partial_{a} S_{R} = - \sqrt{2 M}\, a^{1/2} = - H a^{2}.
\end{equation}
The wave function has a form
\begin{equation}\label{54}
f_{\frac{2}{3}} (a) = \frac{A_{\frac{2}{3}}}{a^{1/4}} \exp \left\{-i \frac{2}{3} \sqrt{2 M}\, a^{3/2} \right \} 
= \frac{A_{\frac{2}{3}}}{a^{1/4}}  e^{-i \frac{2}{3} H a^{3}}.
\end{equation}
This function is the asymptotics of the Airy function at $2 a M \gg 1$. The Airy function is the solution of Eq.~(\ref{12}) with $M(a) = M = const$
for the spatially flat universe (the term $a^{2}$ in Eq.~(\ref{12}) should be omitted). In the domain $a \gg \frac{E}{2 M}$, the
Airy function describes the continuous state with respect to $E$ and can be normalized to a delta-function $\delta(E - E')$ \cite{Lan65}.
The asymptotics (\ref{54}) describes the state with $E = 0$. Therefore, it is convenient to normalize it by the condition
\begin{equation}\label{541}
\langle f_{\frac{2}{3}} (M) | f_{\frac{2}{3}} (M') \rangle = \delta (M - M'),
\end{equation}
where we denote $f_{\frac{2}{3}} (a) \equiv f_{\frac{2}{3}} (a; M)$. In order to calculate the parameters of the universe in the state 
(\ref{54}), the wave function can be normalized to unity
\begin{equation}\label{542}
\langle f_{\frac{2}{3}} (M) | \bar{f}_{\frac{2}{3}} (M) \rangle = 1
\end{equation}
instead of (\ref{541}), where 
\begin{equation}\label{543}
\bar{f}_{\frac{2}{3}} (a; M) = \int_{M - \delta}^{M + \delta} dM' f_{\frac{2}{3}} (a; M')
\end{equation}
is the wave packet with the width $2 \delta \ll 1$, and $\delta > 0$. The amplitude $A_{\frac{2}{3}}$ in Eq.~(\ref{54}) can be found
under the assumption that the probability flux density is conserved in the expanding universe and it equals to the probability flux density
in the radiation-dominated era (\ref{473}). As a result, we obtain
\begin{equation}\label{544}
|A_{\frac{2}{3}}|^{2} = \frac{1}{2 \sqrt{2M}}.
\end{equation}
The wave function (\ref{54}) with the normalizing constant (\ref{544}) no longer has the form of a part of the de Broglie wave.
Its amplitude depends on $a$, and the phase factor contains the momentum (\ref{53}) multiplied by $\frac{2}{3}$,
\begin{equation}\label{545}
f_{\frac{2}{3}} (a) = \left( \frac{1}{2 \sqrt{2M}}\right)^{1/2} \,\frac{1}{a^{1/4}} \exp \left\{i \frac{2}{3} a\, \partial_{a} S_{R}\right \}.
\end{equation}

The case $\alpha = \frac{1}{3}$ in quantum description appears to be special. The wave function has the form (\ref{34}). 
The parameter (\ref{31}) takes the value $\gamma_{\frac{1}{3}} =  - \frac{1}{4}$ and the quantum source function (\ref{16}) is
\begin{equation}\label{55}
Q(a) = - \frac{1}{4 a^{2}},
\end{equation}
so that the energy density (\ref{17}) is negative
\begin{equation}\label{56}
\rho_{Q} = - \frac{1}{4 a^{6}}.
\end{equation}
Then, the condition (\ref{43}) is written as
\begin{equation}\label{57}
\frac{2 M(a)}{a^{3}} + \frac{E}{a^{4}} - \frac{1}{4 a^{6}} - \frac{1}{a^{2}} = \frac{1}{9} \left( \frac{\beta}{a} \right)^{6}.
\end{equation}
Equation (\ref{41}) together with Eq.~(\ref{57}) defines the Hubble expansion rate
\begin{equation}\label{58}
H^{2} =  \frac{1}{9} \left( \frac{\beta}{a} \right)^{6} \quad \mbox{or} \quad H = \frac{1}{3 \tau}.
\end{equation}
This equation describes the spatially flat universe, in which the stiff Zel'dovich matter with the energy density (\ref{56}) dominates. 
From Eq.~(\ref{41}), it follows that for such a universe
\begin{equation}\label{59}
H^{2} =  - \frac{1}{4 a^{6}},
\end{equation}
because the energy densities of a substance and radiation and the curvature term should be neglected. 
This corresponds to the domain of values $a < 1$ (sub-Planck scales).
Equations (\ref{58}) and (\ref{59}) impose restriction on allowed values of the parameter $\beta$
\begin{equation}\label{60}
\frac{\beta^{6}}{9} = - \frac{1}{4}.
\end{equation}
If this condition is satisfied, then it means that in sub-Planck region, where the energy density (\ref{56}) dominates,
the wave function (\ref{34}) is either constant
\begin{equation}\label{61}
f_{\frac{1}{3}} (a) = A_{\frac{1}{3}} \quad \mbox{at} \ \frac{1}{2} = i \frac{\beta^{3}}{3},
\end{equation}
or increases linearly with the increase of $a$,
\begin{equation}\label{62}
f_{\frac{1}{3}} (a) = A_{\frac{1}{3}}\, a \quad \mbox{at} \ \frac{1}{2} = - i \frac{\beta^{3}}{3}.
\end{equation}
In the case of Eq.~(\ref{61}), we have $f_{\frac{1}{3}} (0) = A_{\frac{1}{3}} = const$, i.e. there is a source at the point $a = 0$.
It may indicate that the universe can originate from the initial cosmological singularity point with the finite nucleation rate
$\Gamma  \sim |f_{\frac{1}{3}} (0)|^{2}$. In a more rigorous consideration, it appears that such an origin occurs 
not from the point $a = 0$, but from the whole domain of values of $a \leq \frac{1}{2 \sqrt{E}}$, where the wave function has a form (\ref{61}) 
\cite{Kuz09,Kuz10}. In this domain, there exists the classical trajectory in imaginary time $t = -i \tau + const$, which is the solution of
Eq.~(\ref{41}), where the energy density of a substance and the curvature term are neglected. Near $a \sim 0$, one has
\begin{equation}\label{630}
a = \left(- \frac{3}{2} i \tau \right)^{1/3} \equiv \beta\, \tau^{1/3}.
\end{equation}
From this solution, it follows unambiguously the condition on $\beta$ which chooses the solution (\ref{61}).

In accordance with Eqs.~(\ref{28}), (\ref{42}), (\ref{48}), and (\ref{52}), in both cases $\alpha = \frac{1}{2}$ and $\alpha = \frac{2}{3}$, 
classical theory predicts the existence of the initial cosmological singularity at the point $a = 0$ in the domain of real values of the scale factor.
The solution of Eq.~(\ref{59}) which takes into account the domination of quantum correction (\ref{56}) on scales $a < 1$ demonstrates
that, according to Eq.~(\ref{630}), the initial cosmological singularity lies in the non-physical region of imaginary values of $a$ and it is inaccessible 
from the point of view of general relativity. Moreover, if one takes into account quantum effects in the region $a < 1$, it will allow one to revise
the properties of the universe on sub-Planck scales. Quantum-mechanical description of the universe in semiclassical approximation admits
the possibility of an origin of the universe from the sub-Planck domain.
As is shown in Refs.~\cite{Kuz09,Kuz10}, the origin of the universe is accompanied by a change in space-time topology, so that the geometry
conformal to a unit four-sphere in a five-dimensional Euclidean flat space changes into the geometry conformal to a unit four-hyperboloid
embedded in a five-dimensional Lorentz-signatured flat space. On the boundary, where these two subregions adjoin each other,
there is a jump with change of metric signature \cite{Gib90,Bou98}.

\section{Example}
Let us calculate the probability of the transition of the universe from the state, where
radiation dominates, into the state, in which barotropic fluid in the form of dust is dominant. 
This probability is determined by the expression
\begin{equation}\label{63}
w\left(\mbox{rad.} \rightarrow \mbox{dust}\right) = 
\frac{|\langle f_{\frac{2}{3}} | f_{\frac{1}{2}} \rangle|^{2}}{\langle f_{\frac{1}{2}} | \bar{f}_{\frac{1}{2}} \rangle 
\langle f_{\frac{2}{3}} | \bar{f}_{\frac{2}{3}} \rangle}.
\end{equation}
Using the wave functions (\ref{47}) and (\ref{54}), we have
\begin{equation}\label{71}
|\langle f_{\frac{2}{3}} | f_{\frac{1}{2}} \rangle|^{2} = |A_{\frac{2}{3}}|^{2}\, |A_{\frac{1}{2}}|^{2}\, I(E,M),
\end{equation}
where
\begin{equation}\label{72}
I(E,M) = \left|\int_{0}^{\infty} da\, a^{-1/4} \exp \left\{i \sqrt{E}\, a^{3/2} \left(\frac{1}{\sqrt{a_{c}(E,M)}} - \frac{1}{\sqrt{a}} \right) \right \} \right|^{2},
\end{equation}
and
\begin{equation}\label{73}
a_{c}(E,M) = \frac{9}{8}\,\frac{E}{M}.
\end{equation}
The integration in Eq.~(\ref{72}) can be performed analytically. As a result, $I(E,M)$ will have the form of the sum of terms containing
generalized hypergeometric functions. However, for our illustrative purposes it is sufficient to calculate the integral in Eq.~(\ref{72})
using the method of approximate calculation of overlap integrals of semiclassical 
wave functions mentioned in the Introduction. In the region $a \gg a_{c}$, the exponential in the integrand
oscillates rapidly and its contribution into the integral is exponentially small \cite{Lan65}. 
The main contribution into the integral comes from the region
near $a = a_{c}$, where the exponential is almost unity. In this approximation, we have
\begin{equation}\label{74}
I(E,M) = \frac{3}{\sqrt{2}} \left(\frac{E}{M} \right)^{3/2}.
\end{equation}
Then, taking into account Eqs.~(\ref{472}), (\ref{475}), (\ref{542}), (\ref{544}), and (\ref{74}), we obtain the following simple expression for
the probability (\ref{63})
\begin{equation}\label{64}
w\left(\mbox{rad.} \rightarrow \mbox{dust}\right) = \frac{3}{8}\,\frac{E}{M^{2}}.
\end{equation}
We estimate $a_{c}$ and $w$ using the values $E_{0} = 1.86 \times 10^{118}$ and $M_{0} = 0.92 \times 10^{61}$.
They correspond to the modern values of the energy densities of radiation and matter, 
$\rho_{\gamma}^{0} = 2.61 \times 10^{-10}$ GeV cm$^{-3}$ and 
$\rho_{m}^{0} = \rho_{crit} = 0.48 \times 10^{-5}$ GeV cm$^{-3}$,
and the Hubble length $a_{0} \equiv \frac{c}{H_{0}} = 1.37 \times 10^{28}$ cm taken as a rough estimate of the size 
of the observable universe \footnote{All astrophysical constants and parameters, used here
and below, are taken from Ref.~\cite{PDG}.}.
We have
\begin{equation}\label{65}
a_{c}(E_{0},M_{0}) = 2.27 \times 10^{57} (= 1.69 \times 10^{24} \ \mbox{cm}),
\end{equation}
\begin{equation}\label{66}
w_{0} = 0.83 \times 10^{-4}.
\end{equation}
The redshift $z_{c} = 2.41\, z_{eq}$, where $z_{eq} = 3360$ is the redshift of matter-radiation equality, 
corresponds to the scale factor $a_{c}$ (\ref{65}).
The probability (\ref{66}) has the same order of magnitude as the matter density contrast
$\frac{\Delta \rho_{m}}{\rho_{m}} \sim 10^{-4}$ in the era of matter-radiation equality, when perturbations begin to grow mainly at the expense of
cold dark matter like WIMPs (see, e.g. Ref.~\cite{Row04}).


\begin{thebibliography}{99}
\itemsep -3pt plus 1pt minus 1pt

\bibitem{Lan65} Landau L D and Lifshitz E M, \textit{Quantum Mechanics. Course of Theoretical Physics, Vol. 3},
Pergamon Press, Oxford (1965)

\bibitem{Kuz09} Kuzmichev V E and Kuzmichev V V, 2009 \textit{Acta Phys. Pol. B} \textbf{40} 2877 [arXiv:0905.4142 [gr-qc]]

\bibitem{Kuz10} Kuzmichev V E and Kuzmichev V V, 2010 \textit{Ukr. J. Phys.} \textbf{55} 626

\bibitem{Har83} Hartle J B and Hawking S W, 1983 \textit{Phys. Rev.} \textbf{D28} 2960

\bibitem{Gib90} Gibbons G W and Hartle J B, 1990 \textit{Phys. Rev.} \textbf{D42} 2458

\bibitem{Bou98} Bousso R and Hawking S W, 1998 \textit{Grav. Cosmol. Suppl.} \textbf{4} 28 [gr-qc/9608009]

\bibitem{Ish} Isham C, gr-qc/9510063

\bibitem{Kuch92} Kucha\v{r} K V, in \textit{Proceedings of the 4th Canadian Conference on General Relativity 
and Relativistic Astrophysics}, ed. by G. Kunstatter, D. Vincent and J. Williams, World 
Scientific, Singapore (1992)

\bibitem{Tor} Torre C G, 1992 \textit{Phys. Rev.} \textbf{D46}, R3231

\bibitem{DeW} DeWitt B S, in \textit{Gravitation: An Introduction to Current Research}, ed. by L.~Witten, Wiley, New York (1962); 
1967 \textit{Phys. Rev.} \textbf{160} 1113

\bibitem{Bro} Brown J D and Marolf D, 1996 \textit{Phys. Rev.} \textbf{D53} 1835 [gr-qc/9509026]

\bibitem{Kij} Kijowski J, Sm\'{o}lski A, and G\'{o}rnicka A, 1990 \textit{Phys. Rev.} \textbf{D41} 1875

\bibitem{JKi} Jezierski J and Kijowski J, in \textit{Nonequilibrium Theory and Extremum Principles}, ed. by 
S. Sieniutycz and P. Salamon, Advances of Thermodynamics Vol. 3, Taylor and Francis Publishing Company, 
282-317 (1990) [arXiv:1112.5842 [math-ph]]

\bibitem{SeW} Seliger R L and Whitham G B, 1968 \textit{Proc. Roy. Soc.} \textbf{A305} 1

\bibitem{Sch70} Schutz B F, 1970 \textit{Phys. Rev.} \textbf{D2} 2762, 1971 Phys. Rev. \textbf{D4} 3559

\bibitem{Br92} Brown J D, 1993 \textit{Class. Quant. Grav.} \textbf{10} 1579

\bibitem{Dan} van~Dantzig D, 1939 \textit{Physica} \textbf{6} 693

\bibitem{MTW} Misner C M, Thorne K S, and Wheeler J A, \textit{Gravitation}, Freeman, San Francisco (1973)

\bibitem{Dir58} Dirac P A M, 1958 \textit{Proc. Roy. Soc.} \textbf{A246} 333

\bibitem{ADM} Arnowitt R, Deser S, and Misner C M, in \textit{Gravitation: An Introduction to Current Research}, 
edited by L.~Witten, Wiley, New York (1962) [gr-qc/0405109]

\bibitem{Lan2} Landau L D and Lifshitz E M, \textit{The Classical Theory of Fields. Course of Theoretical Physics, Vol. 2},
Butterworth-Heinemann, Amsterdam (1975)

\bibitem{Kuz08} Kuzmichev V E and Kuzmichev V V, 2008 \textit{Acta Phys. Pol. B} \textbf{39} 979 [arXiv:0712.0464 [gr-qc]]

\bibitem{Kuz13} Kuzmichev V E and Kuzmichev V V, 2013 \textit{Acta Phys. Pol. B} \textbf{44} 2051 [arXiv:1307.2383 [gr-qc]]

\bibitem{Kuz15} Kuzmichev V E and Kuzmichev V V, 2015 \textit{Ukr. J. Phys.} \textbf{60} 664

\bibitem{Dir64} Dirac P A M, \textit{Lectures on Quantum Mechanics}, Belfer Graduate School of Science, Yeshiva University, New York (1964)

\bibitem{Kuc91} Kucha\^{r} K V and Torre C G, 1991 \textit{Phys. Rev.} \textbf{D43} 419

\bibitem{Lin90} Linde A D, \textit{Elamentary Particle Physics and Inflationary Cosmology}, Harwood, Chur (1990)

\bibitem{Kuz02} Kuzmichev V E and Kuzmichev V V, 2002 \textit{Eur. Phys. J.} \textbf{C23} 337 [astro-ph/0111438].

\bibitem{PDG} Olive K A \textit{et al.} (Particle Data Group), 2014 \textit{Chin. Phys. C} \textbf{38} 090001

\bibitem{Kol90} Kolb E W and Turner M S, \textit{The Early Universe}, Addison-Wesley, Redwood City (1990)

\bibitem{Row04} Rowan-Robinson M, \textit{Cosmology}, Clarendon Press, Oxford (2004)

\end{thebibliography}
\end{document}